**Ants3 toolkit: front-end for Geant4 with interactive GUI and Python scripting**

A. Morozov, L.M.S. Margato, G. Canezin and J. Gonzalez

LIP-Coimbra, Departamento de Física, Universidade de Coimbra, Rua Larga, 3004-516 Coimbra, Portugal

E-mail: Andrei.Morozov@lip.pt

# Abstract

Ants3 is a toolkit that serves as a front-end for particle simulations in Geant4 and offers a custom simulator for optical photons. It features a fully interactive Graphical User Interface and an extensive scripting system based on general-purpose scripting languages (Python and JavaScript). Ants3 covers the entire detector simulation/optimization cycle, providing an intuitive approach for configuration of the geometry and simulation conditions, the possibility to automatically distribute workload over local and network resources, and giving a suite of versatile tools based on CERN ROOT for the analysis of the results. The intended application area is the development of new detectors and readout methods. The toolkit has been designed to be user-friendly for those with little experience in simulations and programming.

# 1. Introduction

Monte Carlo simulations of the transport of particles through matter play an important role in the development of new instruments and methods in modern physics, engineering and medicine. Simulations are also invaluable in planning experiments as well as for interpretation of the collected data. Several general purpose toolkits for particle simulations are currently available, including such well-known ones as Geant4 [1-3], MCNPX [4] and Fluka [5].

Geant4 is one of the most broadly used toolkits due to its high flexibility in defining physical models, configuring geometry and organizing scoring. Being open source software makes it much more accessible compared to, for example, MCNPX, which has strict licensing rules. A practical advantage of Geant4 is that it is designed as a c++ library, enabling customisation for specific applications without losing generality. For example, GATE [6] and Geant4-DNA [7] are popular Geant4 extensions for nuclear medicine imaging and modeling of biological damage induced by ionising radiation, respectively.

High flexibility of being a c++ library, however, has its drawbacks. The learning curve of Geant4 is very steep, as even most basic simulations require the definition of multiple classes and overriding virtual functions. Therefore, there are many users, especially at the beginning of their learning cycle, who do not deviate much from the code of the examples distributed with the toolkit.

To address this problem, several “front-end” toolkits were developed, such as GAMOS [8] and TOPAS [9]. They use Geant4 as the simulation engine and handle the configuration of the simulation process without requiring the user to be able to program in c++. Several attempts were also made to provide a Geant4-Python interface, for example, Geant4Py [10].

The existing front-end toolkits have a number of drawbacks that complicate their usage, especially considering beginners. For example, the configuration procedures mostly rely on filling text files using a specific format, thus lacking the interactive character expected of a modern GUI (Graphical User Interface) application. While offering some scripting capabilities (e.g., relying on the so-called “UI commands” of Geant4), they are quite limited in comparison with those offered by general-purpose scripting languages such as Python and JavaScript. As a consequence, defining even somewhat complex geometries can become a non-trivial and time-consuming process. In addition, the scoring system is limited and could even require adding custom c++ code.

In this paper we present Ants3, a new simulation and data analysis toolkit which functions as a front-end for particle simulations in Geant4 and offers a custom photon tracer for optical simulations. Ants3 features a fully interactive GUI as well as an extensive scripting system utilizing standard interpreters of Python and JavaScript. Recompilation is not required for any changes introduced in the configuration, as it is stored in a json-formatted (JavaScript Object Notation) file and the geometry/material information is transferred to Geant4 using a GDML file [11]. As a consequence, modifications in the geometry are immediately visible in the visualisation window.

One of the Ants3 development goals is to shorten as much as possible the time needed for a beginner to start producing meaningful results. Based on this motivation, Ants3 offers several intuitive concepts simplifying geometry definition (e.g., stacks of objects, different types of arrays and instances of a prototype). “Natural” approaches to organize scoring are implemented: for example, a role of a calorimeter can be assigned to any object defined in the geometry or a monitor can be introduced to record statistical information on the passing particles. Flexible options are available for the generation of primary particles and optical photons, and an infrastructure is provided to conduct multi-stage simulations: the output of one stage can be “seamlessly” used to generate primary particles (or optical photons of primary/secondary scintillation) in the next one. Ants3 includes a set of configuration examples which are fully scalable due to the possibility to define geometric parameters using mathematical expressions and user-defined variables, even in the GUI mode.

Ants3 automatically distributes the simulation work over several local processes, and it is possible to additionally use the resources of other computers over the network. The simulation results from different processes are merged and saved to a single file per scorer type. A comprehensive framework is offered for analysis and visualisation of the results, including a customizable event viewer and a large number of statistical tools. The obtained graphs and histograms can be visualised in Ants3 directly using an interface to CERN ROOT [12], included as a library.

The paper discusses first the general structure of the toolkit, then describes the user interfaces, summarizes the geometry and material definition approaches, outlines the particle and optical simulation processes, briefly discusses the result visualisation options and concludes with the implementation details.

# 2. Toolkit structure

The general structure of the toolkit is shown in Figure 1. There are four main components: *Ants3* (configuration hub and user interfaces), *Dispatcher* (work scheduler), *G4Ants3* (particle simulator) and *Lsim* (optical simulator).

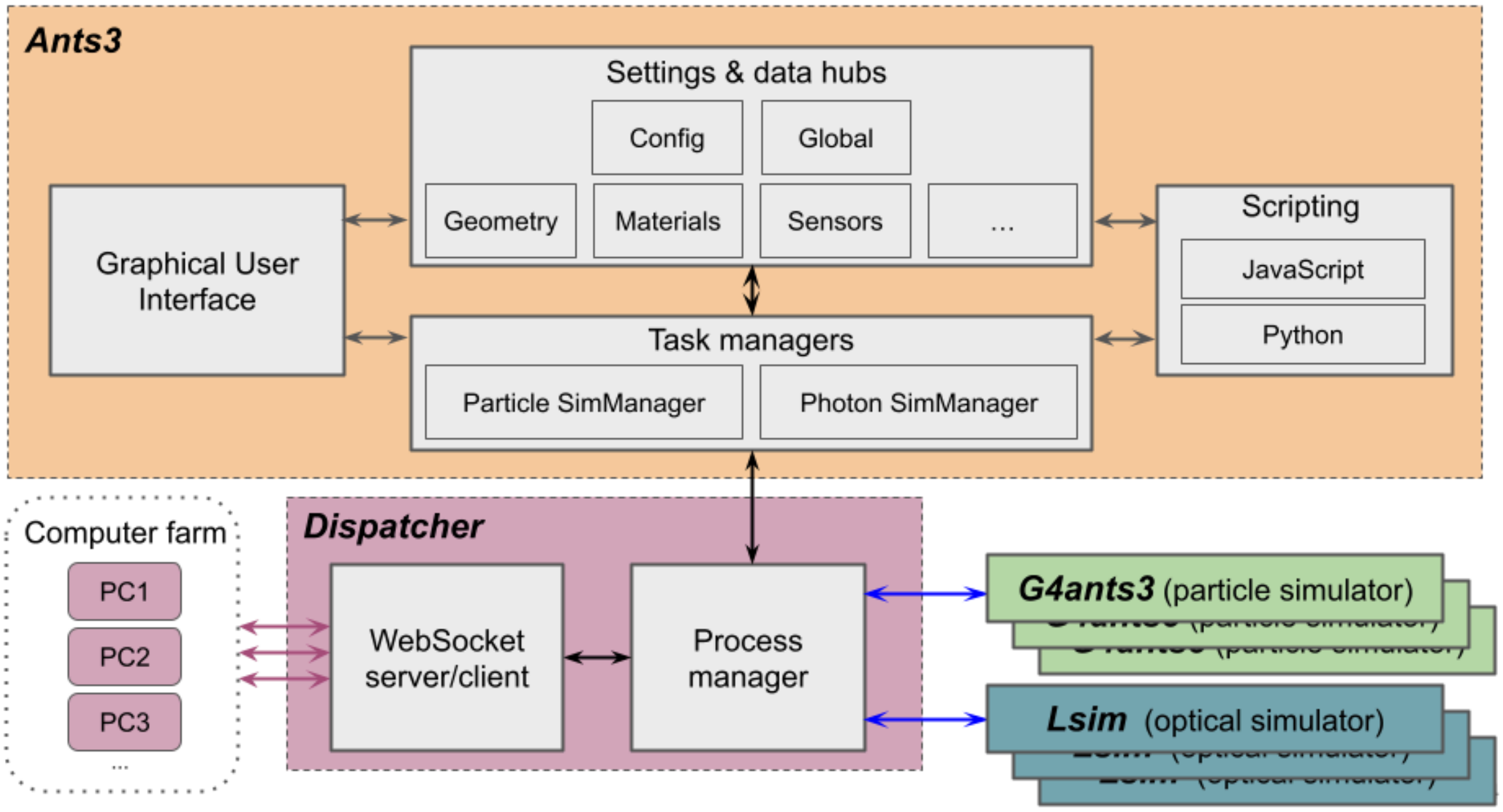

Figure 1. General structure of the Ants3 toolkit.

The *Ants3* component manages all the configuration parameters (e.g., geometry/materials and simulation settings) and provides the user interfaces. The entire configuration can be written to and read from a single json file, enabling flexible modification from the scripting system.

The G4ants3 component is a single-thread Geant4 program which can read the geometry/material configuration from a GDML file and the rest of the simulation parameters from a json config file.

The *Lsim* component is a single-thread c++ program which is a custom optical simulator based on the TGeoManager class of CERN ROOT library. All configuration parameters are read from a json config file.

The *Dispatcher* component manages the distribution of the simulation work by running in parallel several *G4ants3* (or *Lsim*) processes. A part of the workload can also be automatically delegated to other computers by communicating over the WebSocket interface with the Dispatcher instances running on them. Each process is scheduled with a given number of simulation events (the same event cannot be split between several processes).

# 3. User interfaces

## 3.1 Graphical user interface

The graphical user interface of Ants3 is organized as a hierarchical system of windows, each dedicated to a particular task. The main window (see Figure 2) gives access to all secondary windows, and can be used to save / load the entire configuration of the simulation.

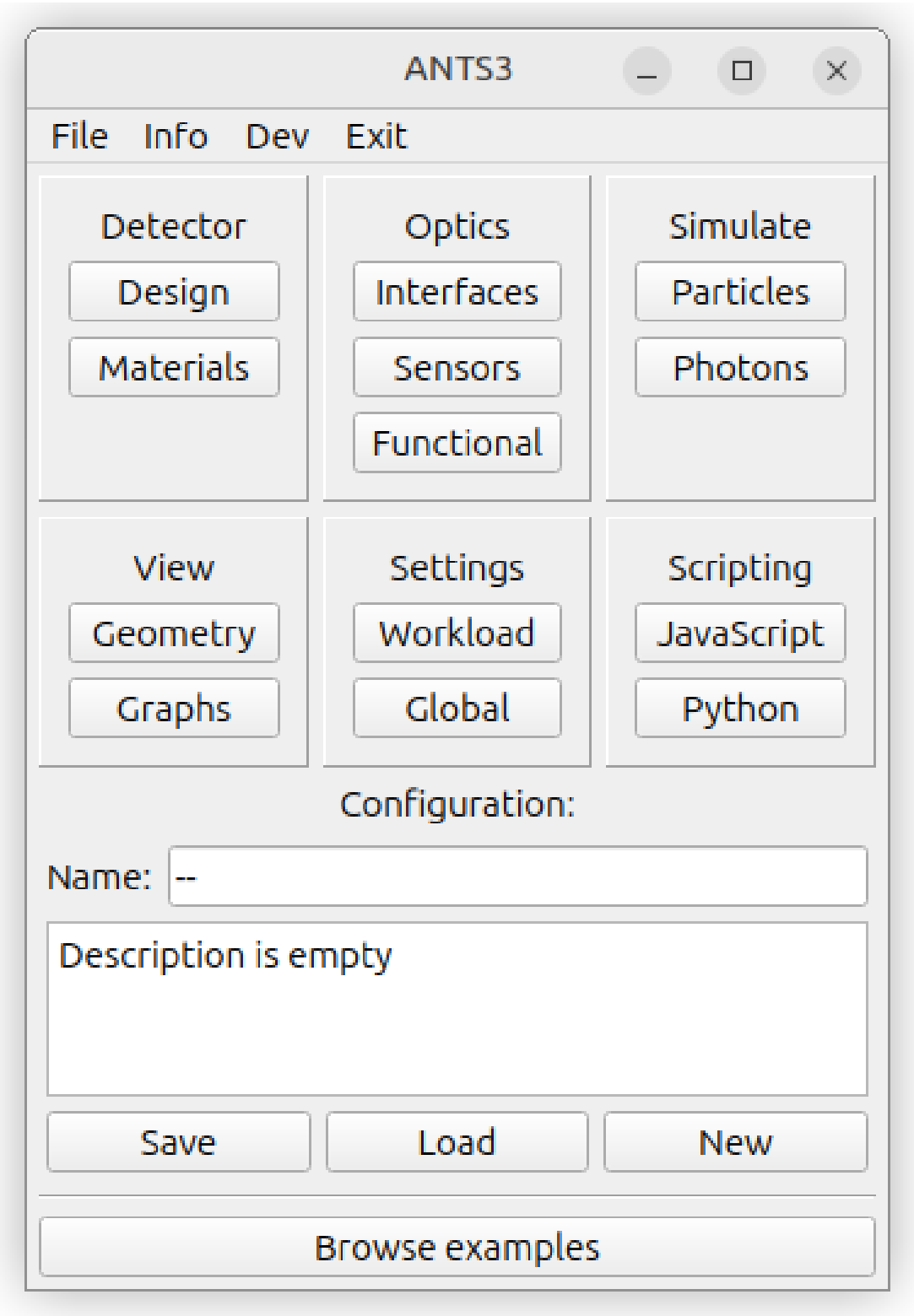

Figure 2. Main window of the Ants3 graphical user interface.

The buttons leading to the second-level windows are divided into six groups. The first one, “Detector”, gives access to the configuration of the geometry and materials. The “Optics” group can be used to configure the optical properties of the material interfaces, light sensors and functional objects (see section 7.2.2). The “Simulate” group leads to the windows dedicated to the particle and optical photon simulations. The “View” group gives access to the geometry visualisation window and the window used to show (and post-process) graphs and histograms.

The “Settings” group is used to configure the distribution of the simulation work between the local and network resources, and to configure the “global” preferences which are stored on the particular computer and not included in the configuration file (e.g., network settings). Finally, the “Scripting” group gives access to the windows used to run JavaScript and Python scripts. Further information on the specific windows is given in the corresponding sections of the paper.

## 3.2 Scripting interface

Ants3 scripting system supports two languages: JavaScript and Python. The Ants3 interface is implemented through custom methods, grouped into *units*. Each unit has a specific set of tasks:

- ***core***: text output, file operations, script execution control
- ***math***: basic mathematical operations extended by custom random generators, fitting and interpolation operations
- ***config***: direct modification of the current configuration (more details below)
- ***geo***: geometry-related tasks
- ***sens***: modification of the optical sensor properties
- ***lsim***: optical simulations
- ***psim***: particle simulations
- ***mini***: access to the ROOT minimizer (more details below)
- ***tracks***: step-by-step analysis of the particle tracking history
- ***partan***: high level analysis of the particle tracking history
- ***farm***: work distribution over the computer farm
- ***graph***, ***hist*** and ***tree***: access to the corresponding ROOT objects
- ***websocket*** and ***webserver***: communication using WebSocket protocol

In GUI mode, the following units are also available:

- ***grwin*** and ***geowin***: access to the graph and geometry visualisation windows
- ***msg***: text output at a standalone window
- ***gui***: toolset for building a simplistic GUI interface (buttons, text input/output fields, etc) reacting to user actions by triggering custom script functions

In both languages, the syntax to call a particular **method** from a specific **unit** is **unit**.**method**(list_of_arguments), for example, psim.simulate() starts a particle simulation run using the current configuration.

As the entire configuration is stored in Ants3 in a json object, all aspects of the configuration can be modified from script by introducing changes in that object. For example, the methods *getConfig* and *setConfig* of the ***config*** unit retrieve and set the entire json object, while methods like *getKeyValue* and *replace* operate on the level of individual json keys. The key argument is provided as a string with the levels of hierarchy separated by a dot: for example, the “ParticleSim.RunSettings.Seed” key refers to the random generator seed used in the particle simulation. Note that to finalize the changes in the configuration, the *updateConfig* method must be called.

The ***mini*** unit can be used to conduct a non-supervised detector optimization (similar to what is described in [13]) or to perform a search of the detector parameters giving the best match between the simulated and experimental results. The user provides the cost function, written in the same scripting language with full access to all Ants3 scripting units. At each minimization step, the function receives an array of values for the free parameters from the minimizer (e.g., Simplex algorithm) and must return the cost value. To compute the cost value, the function can modify the detector configuration, conduct a simulation run and access the results.

A script can be executed in Ants3 with one of the following approaches: 1) using the dedicated scripting window in the GUI, 2) by running the Ants3 executable from the terminal and providing the script file name in the command line and 3) by sending a text message with the script to the WebSocket server (see section 3.3).

### 3.2.1 Running script from GUI

Ants3 GUI has two instances of the Script window, one for JavaScript and another for Python. The Script window has two main areas: the script text and the script output (see Figure 3). Two optional panels can be toggled on the side of the window. The first one displays the current configuration in a tree widget with collapsible items representing json key/value pairs. The second one lists all scripting units and their methods, each with a short description.

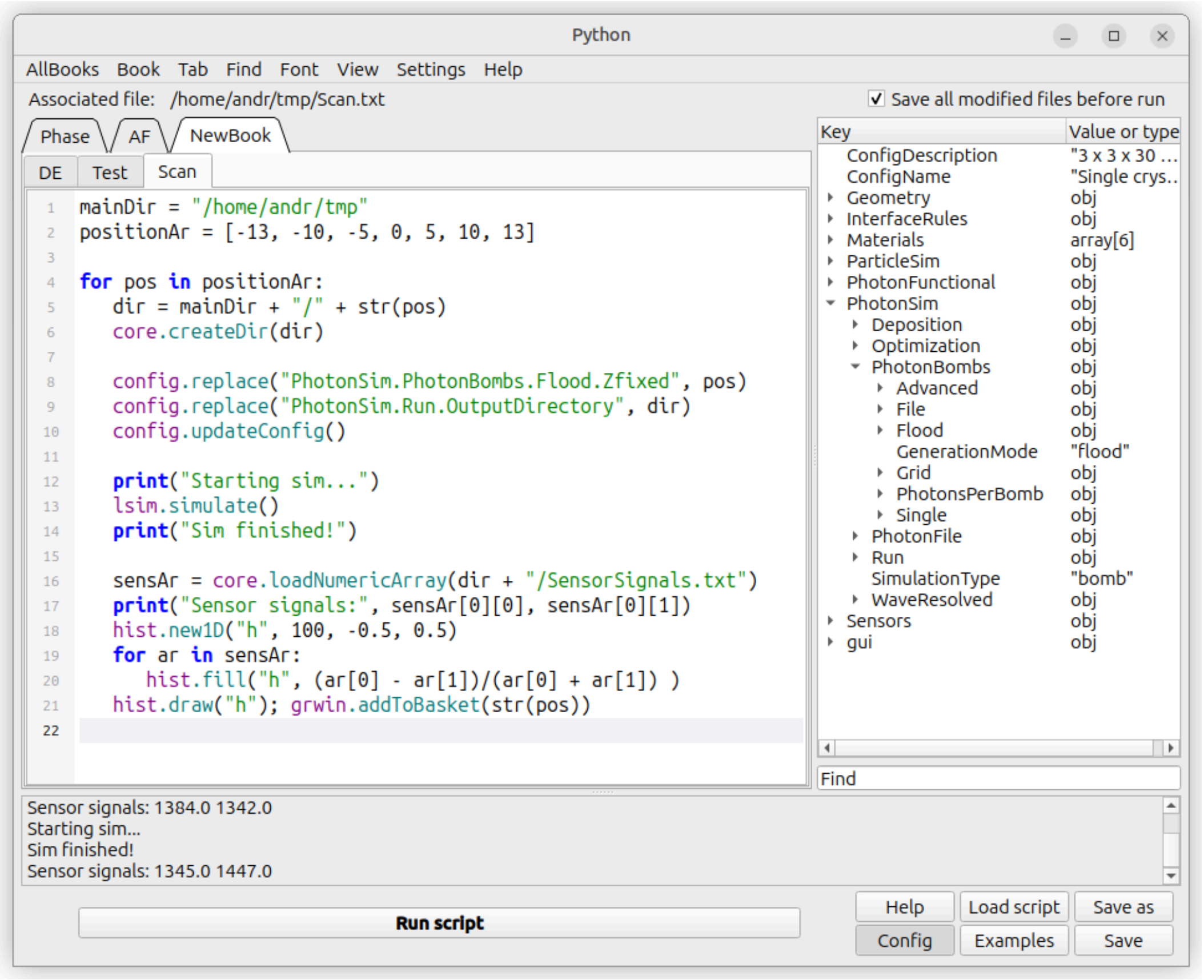

Figure 3. Script window with the configuration explorer panel toggled on.

The script text area allows the user to choose the current script between multiple ones, arranged in *books* and subdivided into *tabs*. The text editor features an autocompleter for the unit methods and user-defined variables. The controls at the bottom of the window are used to start / abort script evaluation, toggle the visibility of the window panels and give access to the script example library.

### 3.2.2 Command line tools for running scripts

Scripts can also be executed from the terminal using the command line interface. The format is *ants3 -j ScriptFileName* for JavaScript, and *ants3 -y ScriptFileName* for Python, where ants3 is the executable and *ScriptFileName* is the name of the text file with the script. Note that when Ants3 is started from the command line, the GUI sub-system is not enabled, and thus the GUI-related scripting units are not available.

## 3.3 WebSocket Server

The third user interface of Ants3 is provided by the WebSocket server. The server can be started from the Global settings window or by running Ants3 in the server mode from the command line ("-s" argument). All incoming text messages to "ws://Address:Port", where Address and Port are the IP address and network port of the server, respectively, are interpreted as JavaScript code with full access to the scripting units of Ants3. Using the infrastructure provided in the ***websocket*** (inbound traffic) and ***webserver*** (outbound traffic) scripting units, it is possible to organize two-way communication with the server over the network or build a custom interface for Ants3 using the localhost.

# 4. Detector geometry

## 4.1 The concept

The geometry definition in Ants3 is based on the TGeoManager class of the CERN ROOT library. This class provides an interface for the geometry configuration and verification, 3D navigation during optical photon tracing and is used in the visualisation of the geometry and tracks. The configured geometry, together with the material properties, can be exported to a GDML file, which is possible to import in Geant4.

The geometry configuration can be defined in Ants3 using several types of objects, all derived from the same base type. Each object has to be positioned inside some other "mother" object (except for the *World* object, which does not have a mother), and can contain an arbitrary number of "daughter" objects itself.

The first type of such objects are physical objects. They have to be assigned with a shape, either one of the elementary shapes defined in ROOT (e.g. box, tube, cone, etc) or a composite shape (union, intersection or subtraction of two or more elementary shapes). The object position and rotation (Euler angles) are defined in respect to the mother object. The object has to be associated with a defined material and can be assigned a single “special role”: e.g. to function as a calorimeter or a light sensor during simulation.

The second type are logical objects. They are introduced to streamline the geometry definition process, and the available options are:

1. *Stack*: the daughter objects of the stack are placed sequentially and without gaps in the Z direction, respecting the order. The daughters can be physical objects or other stacks.
2. *Array*: the complete set of the daughter objects, respecting their relative positions and orientations, defines a template. The template copies are placed forming a rectangular grid pattern, with a given number of times and with a defined step, independently for the X, Y and Z directions. The array daughters can be both physical and any logical objects, and thus it is possible to define nested arrays of arbitrary depth.
3. *Hexagonal array*: similar to (2), but the copies form a hexagonal pattern in the XY plane
4. *Circular array*: similar to (2), but the copies are placed around the Z axis with a defined angular step and radial distance.
5. *Instance*: places a copy of one of the defined *prototypes*. A prototype is a set of objects, physical and/or logical, that has been defined as a template.

All logical objects can have their own position and rotation, which is applied to the generated copies as a common shift/rotation.

It is also possible to define a particle or photon “monitor”: a scoring object of rectangular or round shape with the material of the mother object. The monitor is present in the simulation geometry and records the properties of the particles/photons which enter that object.

Any defined object can be disabled: in this case it is temporarily removed from the geometry until it is enabled. Disabling an object with daughters also disables all of them recursively.

## 4.2 Geometry constants

Any property of an object defined in the geometry, instead of a numeric value, can be given in the form of a mathematical expression using so-called *geometry constants*. For example, if two constants, Size1 and Size2, are defined, a property can be set to “0.5*Size1 - sqrt(Size2) - 25.0”. The syntax of the expression is that of TFormula of ROOT.

A geometry constant is a record with four fields: the constant name, value, optional expression and optional comment. If the expression field is empty, the constant value has to be set directly. The expression allows to automatically compute the constant value based on the values of other constants defined before that one, also using TFormula syntax.

## 4.3 Geometry definition

Geometry can be configured in Ants3 using either the GUI tools or the scripting system. As an advanced option, one can also directly modify the json configuration file.

### 4.3.1 From GUI

The Geometry configurator window can be used to define all geometry-related aspects of the simulation. The configuration process is conducted using three main components of the first tab (*World Tree*) of the window: a geometry tree viewer, an editor for the object selected in the viewer, and a list of the defined geometry constants (see Figure 4).

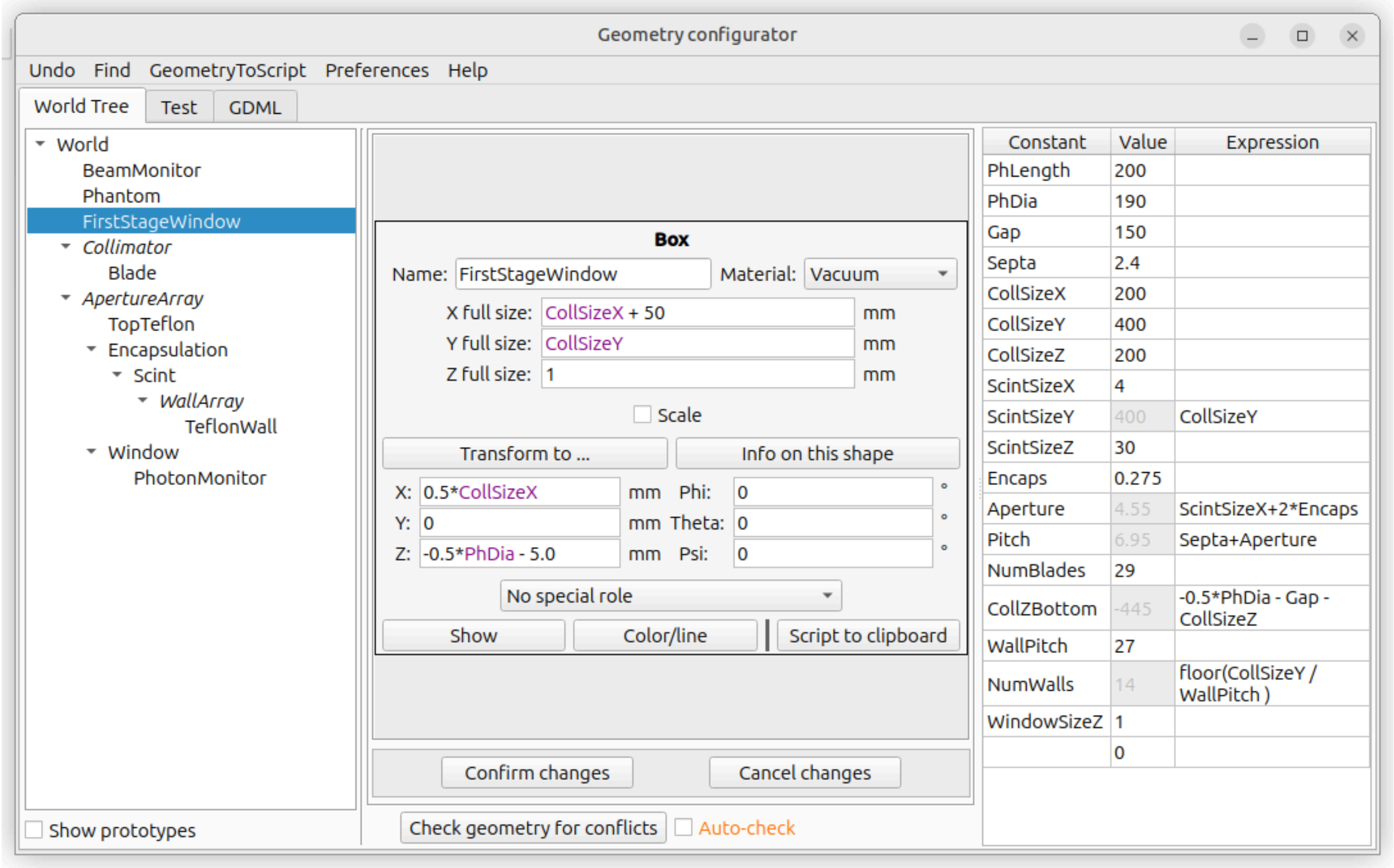

Figure 4. Geometry configurator window.

The geometry viewer is a hierarchical list with each item representing an object defined in the geometry. The objects listed at the same hierarchy level have the same mother volume. Any item can be collapsed to hide its lower hierarchy levels. It is possible to use drag-and-drop to move an object to another mother object or to change the listed order. The context menu of the items gives access to such operations on the object as to add a new daughter object, enable/disable, remove, clone, convert to a prototype, and highlight in the Geometry visualisation window. The “Show prototypes” checkbox at the bottom of the panel enables indication of the defined prototype objects.

When an object is selected in the tree viewer, the middle panel shows an editor for the properties of that object. The list of properties is specific for each object type (or shape for the physical objects). The controls are given for the position and orientation in respect to the mother object and a special role can be selected for physical objects. As discussed in section 4.2, the object properties can be given either as numeric values or mathematical expressions containing defined geometry constants. The auto-completer simplifies the introduction of the constant names, and the highlighter paints the defined constants with a magenta color.

The list of geometry constants shows three fields: the constant name, value and expression. If the expression is not empty, the numeric value is automatically computed and the value field becomes read-only. A new constant can be added at the bottom of the list or above/below a given constant from the context menu. Modifications in the constant values and expressions result in rebuilding the geometry, followed by an update of the visualisation.

The constructed geometry can be checked for conflicts (overlaps and extrusions) using the controls at the bottom of the window. At the second tab (*Test*), a tool is provided to conduct a tracking test through the defined geometry from a selected point in a given direction, listing information on all encountered geometric objects, position of the transitions and the materials.

The GeometryToScript menu item allows to automatically convert the geometry configured in the GUI to a JavaScript or Python script. Geometry constants are converted to script variables and are introduced in the script code. The third tab (*GDML*) provides tools to export the geometry to a GDML file or import geometry from a GDMl file, created, for example, in Geant4.

### 4.3.2 From script

The ***geo*** scripting unit of Ants3 offers all the infrastructure required to define or modify the geometry. There is a specific method to add an object of each particular shape, e.g. a call of *box*(Name, SizeArray, Material, Container, PositionArray, OrientationArray) adds a new box-shaped object named Name of the given material index and the XYZ full sizes as a daughter of the object named Container at the provided local coordinates and orientation.

The logical objects (see section 4.1) are added using the same approach. For example, to form a stack, a new stack object has to be added first to the mother object, and then the content of the stack is filled using the stack object as the container. The same applies to all types of arrays and prototypes/instances.

The scripting unit features methods to remove or enable/disable particular objects, using the object name as an argument. Also, to assign a special role to an object, a scripting method is provided for each available role. The process of defining geometry must be concluded by calling the *updateGeometry* method, which triggers the geometry rebuild process.

### 4.3.3 Using configuration file

As all the geometry-related information is stored in the json configuration file, the geometry can also be defined or changed by directly modifying the content of that file. This method can be used, for example, to create a custom interface for Ants3. The json keys to access the arrays of the defined objects and the geometry constants are Geometry.WorldTree and Geometry.GeoConsts, respectively.

## 4.4 Visualisation

The defined geometry can be drawn at the Geometry visualisation window. There are two available options: a default viewer which uses the infrastructure of the TGeoManager class of ROOT, and a browser based JSROOT [14] (see Figure 5).

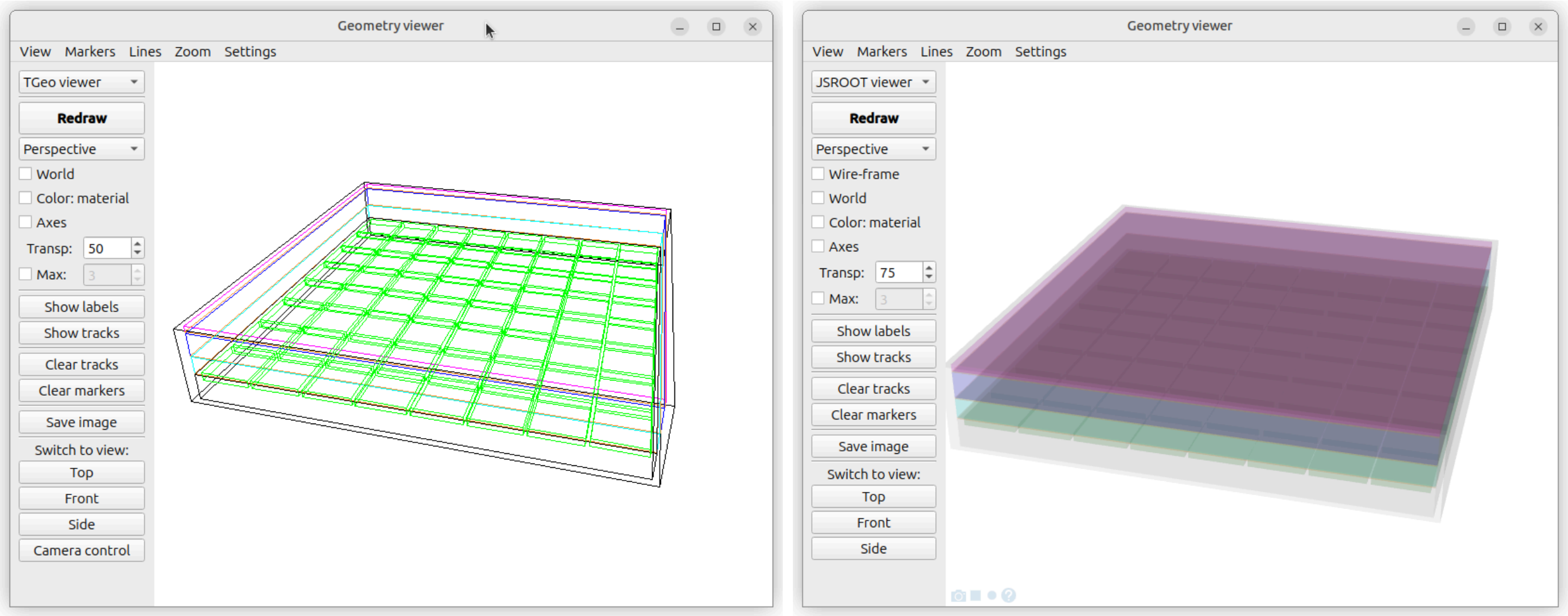

Figure 5. Geometry visualisation window with the default TGeoManager-based viewer (left hand side), and with the JSROOT viewer (right).

Both viewers support controlling the viewing angle and zoom level with the mouse. The default viewer can only operate in the wireframe mode, while JSROOT can also show solids with an adjustable transparency level. The advanced settings of JSROOT enable clipping, precise light source control and exploded view mode.

Controls are provided to show/hide visualisation of the tracks and position markers. “Show labels” button opens a dialog, which can be used to plot different kinds of numeric data on top of the drawn objects, for example, the unique copy indexes of the objects with special roles such as scintillators or light sensors.

# 5. Materials

A material defined in Ants3 is required to have a unique name. All other properties, related to the chemical composition and optical properties, are optional and have to be configured depending on the type and the objectives of the simulation to be performed. This is a natural design choice considering the fact that in a particle simulation, the optical properties are irrelevant, while for an optical simulation, the chemical composition is irrelevant instead.

## 5.1 Material properties for particle simulations

There are two options to define the material composition in Ants3. The first one is to select a material from the Geant4 material database (e.g. G4_H or G3_POLYETHYLENE). The second one is to define a custom composition.

The custom composition is configured with a single string. The following examples illustrate the format of the string:

| | |
|---|---|
| ● C5O2H8 | A regular chemical formula |
| ● C0.5O0.2H0.8 | Real numbers in the formulas are allowed |
| ● {10B:97.0+11B:3.0}4C | A custom isotope composition of an element |
| ● H2O:6 + C2H5OH:3.5 | A mixture defined using molar fractions |
| ● H2O/6 + C2H5OH/3.5 | A mixture defined using weight fractions |
| ● H2O/6 + (NaCl:1 + KCl:2.5)/1.4 | A nested mixture definition |

It is possible to define a composition string using an arbitrary level of nesting with brackets, but the parser of the string can convert molar fractions to weight fractions, but not vice versa.

The density of a material can be defined directly or by declaring the material to be a gas, and providing the pressure and the temperature. It is possible to introduce the mean excitation energy or to delegate the computation of this parameter to Geant4. When a new material is defined, the default properties are those of vacuum.

## 5.2 Material properties for optical simulations

The optical properties of a material in Ants3 can be divided into three groups. The first group is the optical medium properties, such as the refractive index, bulk absorption coefficient, re-emission probability and Rayleigh mean free path. All these properties, except the Rayleigh mean free path, are provided as a value to be used in wavelength-unresolved simulations, and, optionally, as an array of pairs of the wavelength and the corresponding value.

The second group is related to the primary scintillation and includes the emission spectrum, emission rise / decay times, photon yield and intrinsic energy resolution. The rise and decay times can be provided as multiple pairs of the time constant and the corresponding statistical weight.

The third group is related to the secondary scintillation and includes the emission spectrum, energy per electron-ion pair, electron drift speed, transverse and longitudinal electron diffusion coefficients, number of emitted photons per electron and emission decay time.

## 5.3 User interfaces for defining materials

The Material window (see Figure 6) can be used to explore/modify already configured materials, define new ones or remove a material that is not associated with any object in the geometry.

The “Show usage” button highlights in the Geometry visualisation window those objects which are associated with the selected material. A new material can be designed “from scratch” (all properties are set by default to those of vacuum), it can be a clone of an existing one or a material can be loaded from the material library.

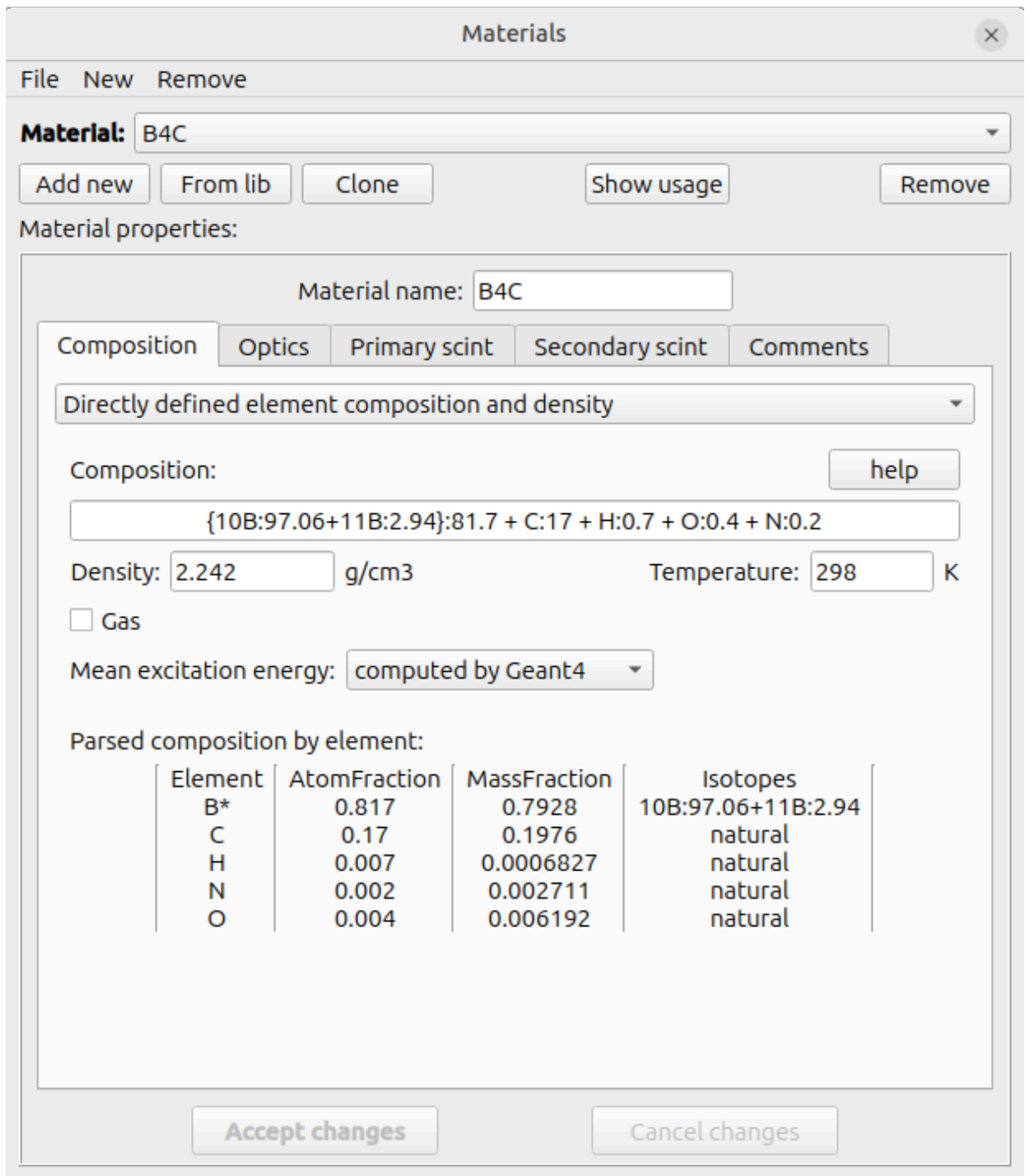

Figure 6. Material window of Ants3.

Access to the material properties from the script is provided by the ***config*** unit, and the path to the json array with the defined materials is “Materials”. After modifying this array, a call of config.updateConfig is required before starting a simulation.

# 6. Particle simulations

This chapter describes the particle simulation cycle of Ants3, including the configuration procedures (physical model, primary particle generator and scorers), the Geant4 simulation phase and the infrastructure available for the analysis of the results.

## 6.1 Configuration

The physics model is configured by choosing the name of one of the reference physics lists of Geant4 (e.g., QGSP_BIC_HP). The maximum allowed tracking step length for charged particles can be defined by providing pairs of the name of an object in the geometry and the associated maximum step value. A list of Geant4 UI commands can also be introduced to be executed before starting the Geant4 run. For example, the command “/run/setCut 0.7 mm” (the default content of that list) sets the production cut-off to 0.7 mm.

These settings can be configured in the Particle simulation window (*Settings* tab) or from script by modifying the value of the ParticleSim.Geant4Settings key in the configuration object.

## 6.2 Primaries

There are two options to generate primary particles in Ants3 simulations: by configuring the "particle sources" or by providing a file listing individual particles.

### 6.2.1 Sources

A particle source is a software routine that generates a primary particle (or a number of primary particles) based on the user-defined parameters. In Ants3, a source has the following properties: a name, particle composition, position, shape (with orientation), as well as parameters defining the angular and temporal properties.

The particle name should identify the particle in the configured physics list of Geant4: for example, "e-", "gamma" or "Na22". The information on the initial energy can be provided as a fixed value or a custom distribution. Each particle has to be declared either as *independent* or *linked*. One of the independent particles is selected each time the generation routine of the source is triggered based on the statistical weights defined for each of these particles.

A linked particle has to be associated ("linked") with a "mother" particle, and it is generated with a given probability depending on the generation outcome for the mother particle. The trigger can be the generation of the mother particle or the fact that it was not generated. The latter option allows configuring alternative decay paths. It is possible to define an arbitrary number of linked particles and a particle can be linked to another linked particle, thus enabling configuration of particle cascades. A possibility is also given to generate a pair of identical particles with the opposite directions (e.g., for simulating annihilation gammas).

The source shape defines a region in space where the emission positions are uniformly generated. The available options are a point, line, rectangle, round, box or cylinder. An arbitrary shape can also be configured by defining a limiting material: in this case, the positions which are not inside an object of the specified material are rejected during generation.

A source can be isotropic or the emission direction can be either fixed or sampled from a custom distribution. A cut-off angle can be introduced to reject directions outside of a given cone. For the "fixed" option it is also possible to introduce Gaussian divergence.

The temporal properties are controlled by two main parameters: the time offset and the spread. The time offset can be a fixed value, a value sampled from a custom distribution or one derived from the event index. The event index option can be used to simulate regular bunches of particles by defining the start time and the bunch period. In this case, the offset is set to the start time plus the event index multiplied by the bunch period. The spread options are none, Gaussian, uniform and exponential. The offset and spread values are generated independently, and the particle time is set to their sum.

An arbitrary number of sources can be defined in a particle simulation. For each simulation event, one source is randomly selected based on the provided statistical weights. It is also possible to trigger the same source multiple times per event. The number of times can be fixed or sampled from a Poisson distribution with a given mean value.

The sources can be defined in the Particle simulation window using the particle source dialog (see Figure 7) or from script by accessing the array of sources in the configuration json using the ParticleSim.GenerationFromSources key.

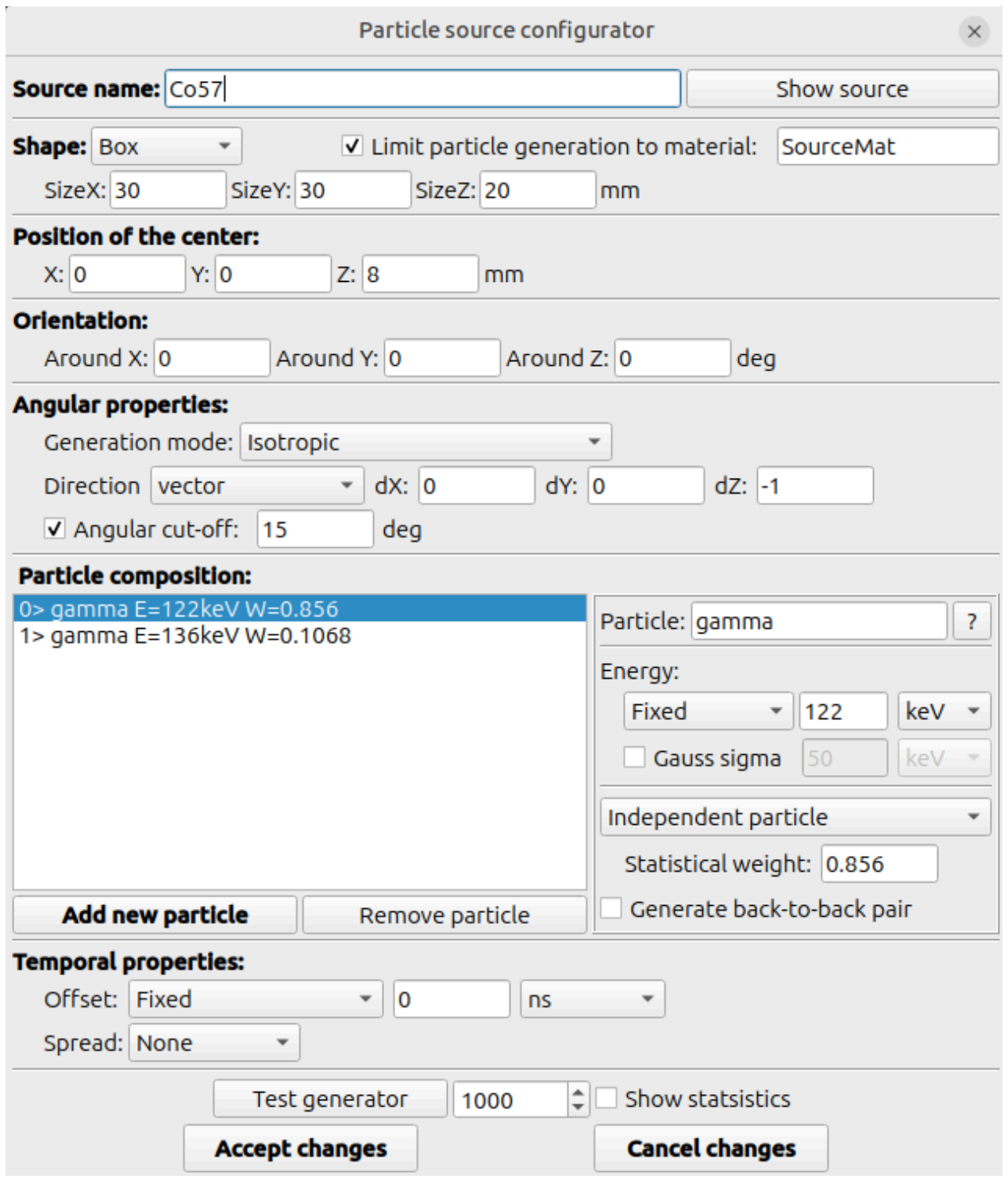

Figure 7. Particle source dialog.

Note that the source dialog offers an infrastructure to check the configured source ("Test generator" button). The generator is called a given number of times, collecting the distributions of the particle composition, energy and time of emission. The generated positions and directions are shown in the Geometry visualisation window.

## 6.2.2 List file-based

The primary particles can also be loaded from a file (ascii or binary). The file lists individual particle records, separated by event markers. Each particle record contains the particle name (as defined in Geant4), kinetic energy, position, direction unit vector and the time stamp. Detailed information on the format is provided in the GUI (Particle simulation window).

## 6.3 Scoring

Scoring methods for particle simulations in Ants3 range from the “brute-force” logging of the entire tracking history of all particles to ones based on collecting statistical information on the selected particles or logging such data as energy deposition in the specific objects. Several alternative methods of obtaining most types of data are offered, and the choice of the approach can be made considering how detailed and versatile information is obtained versus the required resources (simulation time and the output size).

### 6.3.1 Tracking history

The most versatile scoring approach is storing the entire tracking history in a file. A record is created for each tracked particle, containing the particle name, name of the process which resulted in the generation of that particle, and two lists: one containing all tracking steps and the other with the records of all secondary particles. For each step, the particle position, kinetic and deposited energies, time, and indexes of the secondary particles generated in that step are recorded. The drawbacks of using this scoring method are a strong impact on the simulation speed and potentially very large size of the output files for complex geometries, as every step, including the transportations from one geometric object to another, is recorded.

### 6.3.2 Monitors

As discussed in section 4.1, Ants3 geometry can contain *monitor* objects. A monitor records the distributions of the properties for the particles which enter that object, such as the kinetic energy, time, position and angle of incidence. The position and angle are in the local coordinates of the object. A monitor can be sensitive to all particles or just a specific one, and it is possible to limit sensitivity to either primary or secondary particles, as well as to particles with or without the prior interaction which could change their direction. A monitor can be configured to record particles entering only from a specific direction. There is also an option to stop the tracking for particles after recording their data.

If a monitor is hosted by any hierarchy of logical objects that create copies, all monitor copies record data independently. Every monitor, including the copies, receives a unique index in the geometry, enabling its identification in the output.

### 6.3.3 Calorimeters

Objects defined in the geometry can be assigned with the special role of *calorimeter*. The corresponding scorer records information related to the energy deposition inside the boundaries of that object. There are three data collection modes. The first one is “Energy per event”: the calorimeter records the distribution of the total energy deposited per event. The other two modes, “3D energy” and “3D dose” are used to record the three-dimensional spatial distribution of the deposited energy and dose, respectively, cumulatively over all events. The binning of the corresponding histogram is defined by the user, and the dose is computed as the ratio of the deposited energy and the mass of the corresponding voxel assuming it is entirely made of the material at the position where the deposition has happened. A calorimeter can be configured to be “composite”: in this case, energy deposition in any daughter object (recursively) with no special role, triggers the scoring procedure of the calorimeter. Similarly to monitors, all calorimeters, including their copies, record data independently.

### 6.3.4 Particle analyzers

A *particle analyzer* is another special role that can be assigned to an object. The corresponding scorer records information on the particles entering that object or, alternatively, on the particles created inside. The recorded data, by particle name, are the number of particles and their energy distributions. Each analyzer can be configured to operate in the *individual* or *common* scoring mode. In the individual mode, if multiple copies of that analyzer are present in the geometry, each copy records data independently (similar to the monitors and calorimeters). In the common mode, all copies of the same analyzer share data.

### 6.3.5 Energy deposition

Another scoring option of Ants3 is the energy deposition logging. Whenever a particle tracking step results in energy deposition inside one of the objects on the user-defined list, a record containing the particle name, deposited energy, position, index of the material at this position, time and the object copy index is stored in the output file. The list of objects is defined by the user with an option to include all objects with the special role of *scintillator*. Note that all copies of a scintillator object receive unique copy indexes, even if they appear in nested arrays, which is not enforced for objects without a special role. The output file also contains event start markers.

### 6.3.6 Exiting particles

Particles that exit a specified object can be logged in a file. The records are stored using the same format as the one defined for the list files with the primary particle (section 6.2.2), thus giving a straightforward way to organize multi-stage simulations. It is also possible to limit logging only to particles with the timestamp within a given interval.

## 6.4 Running the simulation

After the geometry, materials, primary sources, physical model and the scoring options are configured, the final step before running a simulation is to set the number of events, the seed of the random generator and the output options. These options include the output directory and the enable/disable flag for each scorer type (e.g., to generate the tracking history or not).

The first approach to start a simulation run is to use GUI controls at the *Simulation* tab of the Particle simulation window. The second one is to call the *simulate* method of the ***psim*** scripting unit. For both approaches, Ants3 distributes the workload automatically over the configured number of local processes and those on the configured nodes of the farm (see section 2). The output files from all processes are automatically merged and the resulting files, one per enabled scorer type, are saved in the configured output directory.

There is also an advanced approach, targeting simulations on a computer cluster. It is possible to run g4ants3 executable directly from the terminal with the name of the configuration json file provided as an argument.

## 6.5 Analysis tools

Ants3 offers a comprehensive system of analysis tools, both in the GUI (*Results* tab of the Particle simulation window) and from the scripting interface.

### 6.5.1 Tracking history-based tools

The first group of analysis tools is dedicated to the processing of the particle tracking history. The raw data extracted from Geant4 is usually convoluted due to the “reverse” tracking order implemented in Geant4: a Last-In / First-Out container is used to store the list of particles to be tracked, so the last added secondary particle is tracked first. Therefore, in Ants3, the tracking data are first loaded to a tree container, in which each entry is a particle track with sub-entries hosting secondary tracks generated by that particle.

#### 6.5.1.1 Visualisation of tracks

The *Tracks* tab of the GUI contains controls to visualise tracks. It is possible to show all tracks, only tracks of the primaries or only those of the secondaries. Options are given to suppress tracks of a configurable list of particles or only show tracks for particles from another list. Note that the track visualisation options, such as color, thickness and line style, can be configured at the *Settings* top-level tab of the window, both the default properties and, optionally, individually for each particle by name.

#### 6.5.1.2 Event viewer

The *Event viewer* tab can be used to explore an event in the tracking history visualised as a tree (see Figure 8). The viewer lists the primary particles (top level items) and, as collapsible sub-items, the records of each tracking step of the corresponding particle. The step record shows the name of the process which defined that step (as generated by Geant4 with the exception of “C” - creation, “T” - transportation, and “O” - exiting the defined geometry) followed by a configurable list of the step-related information (e.g., position, time, material etc, see Figure 8). If a secondary was generated at a given step, its record appears as a sub-item of that step item. The tracks of the current event are shown in the Geometry visualisation window, and several additional options (e.g. to center the view at the specific step) are accessible through the context menu of the corresponding step item.

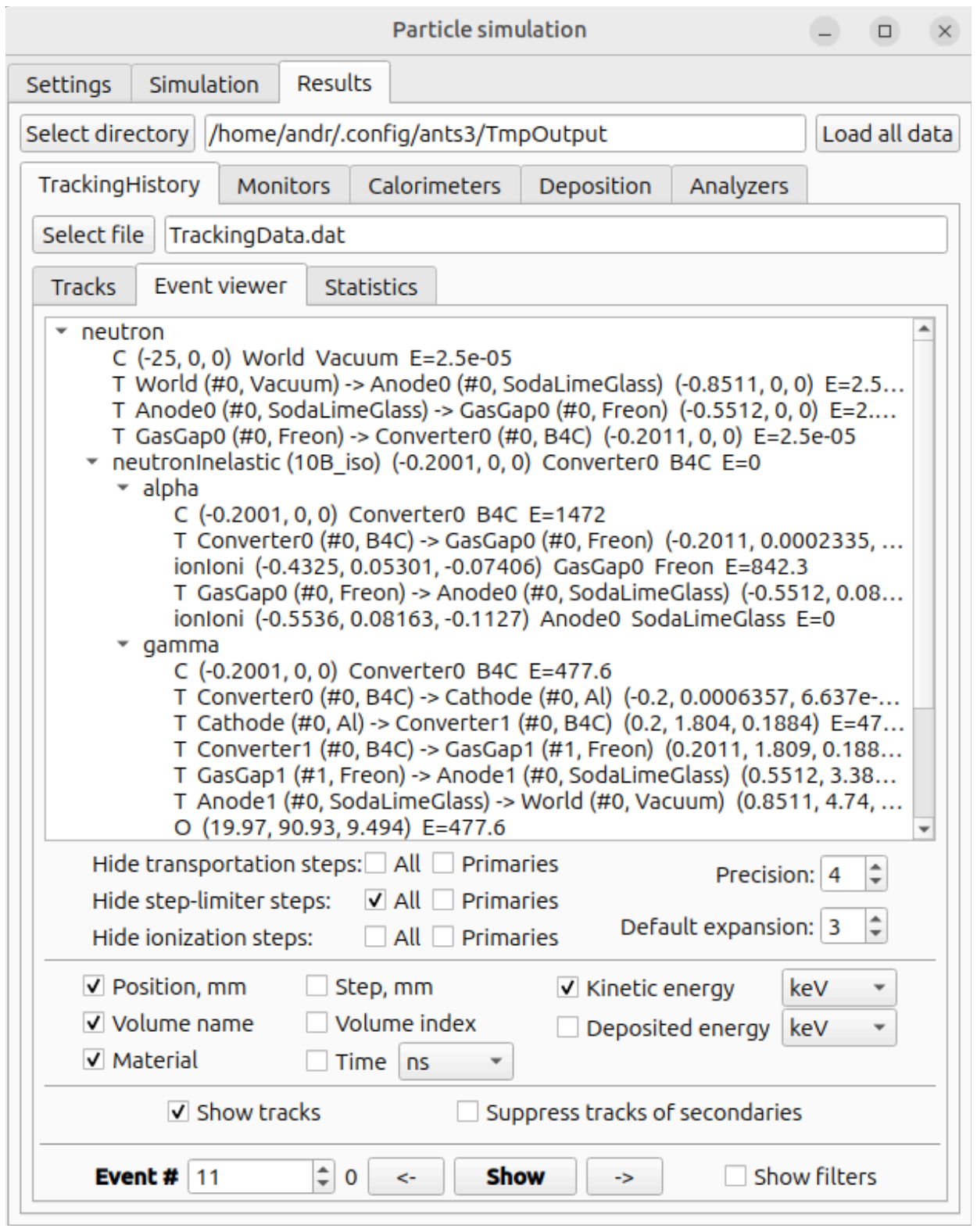

Figure 8. Event viewer

The event index can be introduced directly or it is possible to advance to the next event satisfying a set of filters, limiting, for example, the process, particle or object in the geometry.

The tracking history can also be accessed through the ***tracks*** scripting unit. After loading the history from a file, one can choose the current particle track (by selecting the index of the event and of the primary) and request the track record. It is possible to follow step by step the track, and switch to (and out of) one of the generated secondaries. A detailed description for each method can be found in the GUI (Script window).

#### 6.5.1.3 Statistics

Ants3 offers a set of tools to facilitate statistical analysis of the tracking history data. They are intended to cover the most common scenarios of the data analysis thus often eliminating the need to create custom scripts “scanning” the history. There are two groups of these tools: the first one is dedicated to collecting information from the bulk of the objects defined in the geometry, while the second one targets transition between the objects.

The tools of the first group (*In volume* tab in the GUI) collect information such as statistics on the particles and processes appearing in the tracking history, as well as the distributions of the travelled distances or deposited energies. It is possible to limit the collection of data to a specific combination of the material, object name and object copy index.

The tools of the second group (*Transitions* tab in the GUI, see Figure 9) operate differently. They collect one- and two-dimensional distributions of the quantities which can be represented as a mathematical expression using the following parameters: the global coordinates (X, Y and Z), direction unit vector (Vx, Vy and Vz components), energy and time. For example, a valid search request is to build a distribution of “Energy*1000” versus “sqrt(X*X + Y*Y)”. A limiting condition can be added using the same parameters, for example, “Time < 0.5 || Time > 1.5” (syntax of TFormula of ROOT). To identify the specific transitions, it is possible to configure a given combination of the exit and entrance objects by material, name and copy index. One can use “Limit to escaping” or “Only created” flags to collect information on the particles escaping the World object, or on the generated particles before tracking.

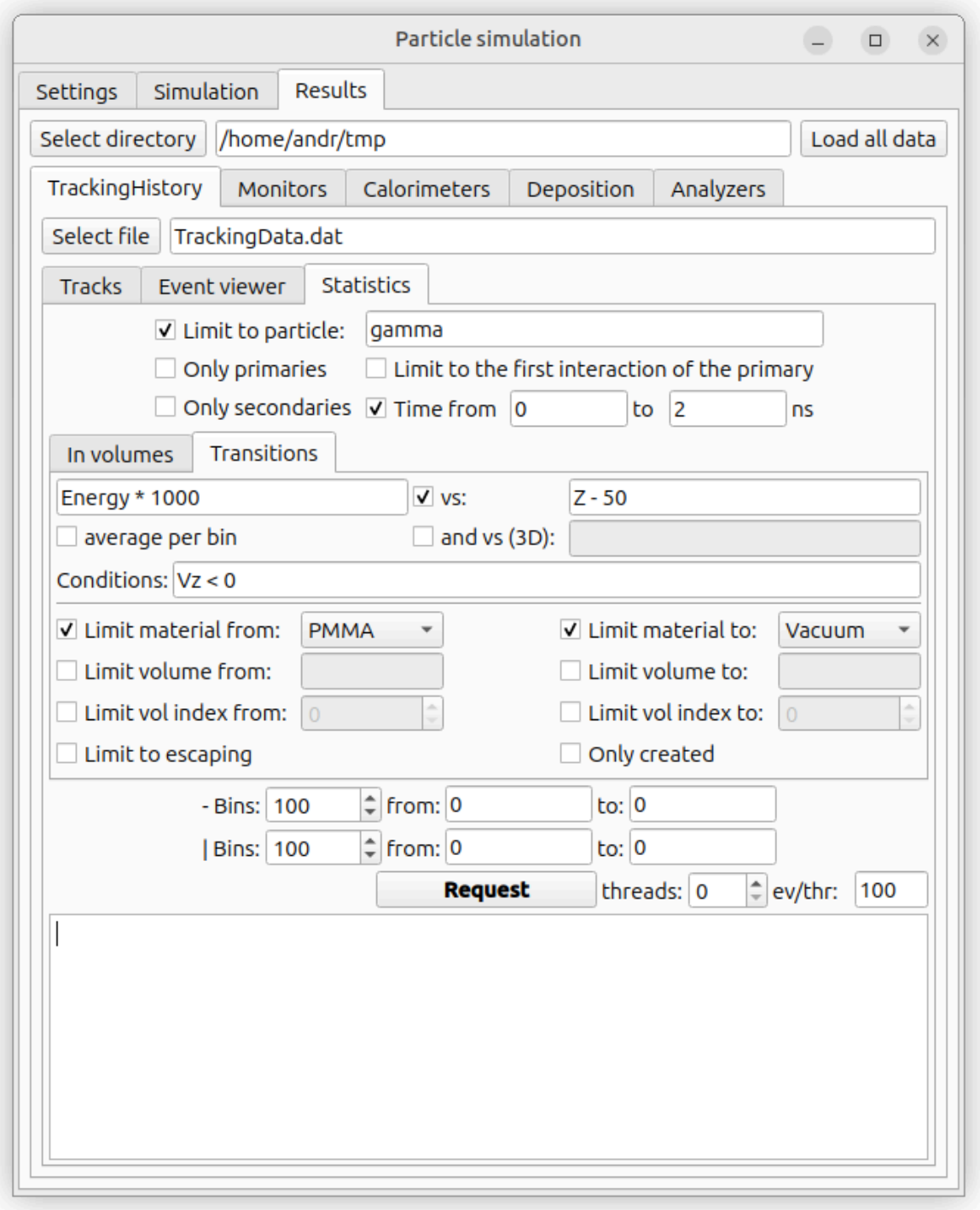

Figure 9. Tools for extracting statistical data from the tracking history.

The tools of both groups can also limit the search to a specific particle or particle property (e.g. to primary or secondary particles) and to a given time window.

Both types of tools can also be used from script (***partan*** scripting unit). The procedure is to first load the tracking data file, then configure the search criteria on the particles and objects, and finally, either use one of the in-volume search methods (e.g., *findProcesses*) or, for the transitions, the *findOnBorder* method.

### 6.5.2 Monitors

The *Monitors* tab gives access to the data collected by the monitors. After selecting one of the defined monitors by its unique index, an interface is shown to request a plot of the distribution of the hit position, time, angle of incidence and energy. The unique indexes of the monitors can be shown in the Geometry visualisation window ("Show labels" button), and a list with the coordinates of the monitor centers, by the monitor index, can be requested in script (***psim*** unit).

If several monitors are defined, it is also possible to plot a graph with the number of monitor hits versus the monitor index and the global time distribution of the hits over all monitors. The latter is only an approximation if the time binning of different monitors is not identical.

The access to the monitor data from script is provided by the ***psim*** unit. After the data file is loaded (*loadMonitorData* method), a specific distribution can be requested by calling the corresponding methods (e.g., *getMonitorEnergy*) and giving the monitor index as the argument.

### 6.5.3 Calorimeters

The *Calorimeters* tab gives access to the data collected by the calorimeters. After one of the calorimeters defined in the geometry is selected by its unique index, the available GUI controls depend on the configured operation mode of that calorimeter. For the "Energy per event" mode, the only option is to plot the distribution of the deposited energy per event. For the "Deposited energy" mode, there are two options: to plot a "projection" or a "slice". For the "Dose" mode, only plotting of a slice is available. Both "projection" and the "slice" allow to indicate one axis or two axes along which the data will be plotted. For "slices", additional controls are used to select the slice position. The calorimeter data can also be accessed from script, and the procedure is similar to that for the monitors.

### 6.5.4 Deposition

The controls at the *Deposition* tab can be used to perform a high-level analysis of a file with the energy deposition data. The report prepared by the analyzer lists the total number of events, the energy deposition statistics separately for each particle type, minimum and maximum deposited energies per event, the time interval spanned by the depositions and the spatial range where depositions were recorded.

The energy deposition data can be imported in script by using the infrastructure for loading three-dimensional arrays of the ***core*** unit (e.g. *load3DArray*). The array dimensions of the imported data are the events, deposition records of one event, and the fields in one deposition record. A detailed information on the format is provided in the GUI.

### 6.5.5 Particle analyzers

The *Analyzers* tab gives access to the data collected by the particle analyzers. Similar to the other scoring modalities, the first step is to select an analyzer. Note that if for a given analyzer the "Single instance for all copies" option is selected, only one record is generated for that analyzer. The report shows the number of occurrences for each particle and the corresponding mean kinetic energy. It is also possible to request a plot of the energy spectrum for a selected particle. The access to the particle analyzer data from script is also provided by the ***psim*** unit.

# 7 Optical simulations

This section describes the design of the custom tracer of Ants3 for optical photons, the optical simulation cycle and the infrastructure available for the analysis of the results. Note that in this chapter we use the term “photon” exclusively for optical photons.

## 7.1 The simulator

Ants3 uses a custom photon tracer, based on the 3D navigator of the TGeoManager class of the ROOT toolkit. A simulation event consists of generating and tracing, one by one, a certain number of photons. The life cycle of a photon starts from assigning it an initial position, direction, time and wavelength. The tracing procedure assumes straight line propagation of photons between interactions, which can be in-bulk (e.g. absorption or Rayleigh scattering) or on the interface between two objects. By default, the interactions on the interface are processed according to the Fresnel and Snell laws and assuming an ideally flat surface. It is possible to introduce custom interface rules (section 7.2.1) to directly define the reflection, scattering and absorption probabilities or to activate one of the available interface models (e.g. rough surface or wavelength shifter). Custom tracing rules can also be assigned to any physical object defined in the geometry by setting its special role to *Functional model* (for example, a thin lens, see section 7.2.2). For all optical processes, it is assumed that photons are unpolarized. The scoring can be conducted using two main approaches. The first one is to assign the special role of *Optical sensor* to some objects in the geometry (section 7.4.1), and the second one is to introduce optical monitors (section 7.4.2).

Ants3 can conduct simulations in wavelength-resolved and wavelength-unresolved modes. For a wavelength-resolved simulation, a binned wavelength range has to be configured by setting the minimum and maximum wavelengths and the step. All optical properties, provided by the user as arrays of wavelength/value pairs, are automatically interpolated to the same regular grid of wavelengths in that range. Each generated photon receives a “waveindex” - the node index of that grid. During wavelength-unresolved simulations all photons have the waveindex value of -1, and the “effective” values of the optical properties are used, which also have to be configured by the user. Note that even in wavelength-resolved mode, a photon can be attributed a waveindex of -1 if it was, for example, emitted from a material without a defined emission spectrum.

Another important note is that in the case when only an effective value is defined for a given optical property, this value is used in wavelength-resolved simulations. Not requiring to provide the wavelength-resolved properties for all parameters might seem counter-intuitive, but this choice simplifies the definition of materials in case only particle simulations are intended or a given optical property does not change with the wavelength or is not relevant for the simulations.

## 7.2 Configuration

### 7.2.1 Custom interface rules

A custom interface rule is a model that is triggered when a photon arrives at a specific interface between two objects in the geometry. A concrete interface is specified as a transition from one defined material to another by configuring a pair of materials, or, alternatively, as a transition from one object to another by providing the object names.

Ants3 offers several interface rules, for example, the "Simplistic" rule allows to configure the probabilities of the specular reflection, absorption and Lambertian scattering. The "Unified" rule assigns the surface model implemented in Geant4 with the same name. The "RoughSurface" rule can be used to simulate interaction with surfaces of significant roughness: one can select a Glisur or Unified models of Geant4, or a microfacet model with a configurable custom distribution of the microfacet normals. The full list of the available interface rules and their descriptions can be found in the GUI (Optical interface window).

If for a certain interface both a material-to-material and object-to-object rules are defined, the object-to-object rule is only triggered. It is recommended to avoid assigning a large number of object-to-object rules, as it can have a negative impact on the simulation speed.

The interface rules can be configured in the Optical interface window. After selecting a model and configuring its parameters, the "Interface rule tester" dialog can be opened. The dialog offers several tools to characterize the effect of this interface on photon tracing, e.g., by computing the absorption/reflection/scattering fractions as a function of the angle of incidence. The rules can also be configured from script by accessing the "InterfaceRules" key of the config json. Two sub-keys, "MaterialRules" and "VolumeRules", address the arrays of the defined material-to-material and object-to-object rules, respectively.

### 7.2.2 Functional models

Assigning the special role of *Functional model* to an object defined in the geometry results in replacing the standard photon tracing procedure inside that object with a custom one. Note that photons interact in the standard way with the interface of that object, and only after transmission inside the object, the custom procedure is triggered.

At the moment of writing this paper, three models are implemented: an optical filter (gray or bandpass), a thin lens (focal length can be given as a function of the wavelength) and an optical fiber. The optical fiber model allows to "teleport" photons between two objects and the fiber acceptance angle can be configured as a function of wavelength. The Ants3 development plans include the implementation of additional models, for example, introducing the point spread function formalism.

The properties of a functional model can be reconfigured independently for all copies of the object with that role. The controls provided in the GUI (Functional model window) or the *configurePhotonFunctional* method of the ***geo*** scripting unit can be used for that purpose.

More details on using functional models can be found in the configuration examples ThinLens and WSFneutron, illustrating the use of the thin lens and optical fiber model, respectively.

### 7.2.3 Optimization options

It is possible to configure the limit on the total number of steps a photon can have during tracing. The limit is required as there are conditions in which a photon can be trapped in an infinite tracing cycle, due to, for example, full internal reflection. By default, the maximum number of steps is 500, and it is recommended to choose the value based, for example, on the distribution of the number of steps photons have before detection by a sensor (see section 7.6.5).

The second option allows to conduct a preliminary detection test for each photon before tracing, and skip tracing if the test fails. Activating this option can significantly speed-up simulations for configurations with optical sensors of low detection efficiencies. The procedure is the following: a random number is generated and tested against the pre-computed maximum detection efficiency over all sensors, including all possible modifiers due to, e.g., angle of incidence. The photon is traced only if the test is passed, and the standard detection procedure is applied (using the already generated random number) if the photon hits a sensor.

## 7.3 Photon generators

Ants3 offers three different approaches to generate photons in optical simulations. The first one is to specify the position and number of photons directly (or sample them from a distribution). The second is to generate photons based on the scintillation properties of a material and the energy deposition information provided by a particle simulation. The third is to load information for each photon individually from a list file. The generators can be configured in the Particle simulation window or from script by accessing the config json with the “PhotonSim” key.

### 7.3.1 Photon bombs

“Photon bomb” refers to the direct generation of a certain number of photons at a given position. Ants3 offers several options both for choosing the number of photons per bomb and for the generation of the bomb positions.

The number of photons per bomb can be a constant value or a value sampled from the Poisson, Gaussian, Uniform or a custom distribution. The generation options for the bomb position are: a directly defined single one; a regular grid (up to three dimensions); a position chosen uniformly in a defined region (2D or 3D); or positions loaded from a list file.

By default, the photon directions are generated assuming isotropic distribution. For wavelength resolved simulations, the photon wavelength is sampled from the primary scintillation spectrum defined for the material at the position of the emission. Similarly, the photon timestamp is generated based on the rise and decay times configured for that material.

There are also advanced options that allow to override the default emission properties. It is possible to assign a fixed direction to all photons, set a fixed wavelength and a timestamp. Also, for all position generation options except “single”, it is possible to skip those positions which are either not inside an object with the configured name or when the material at that position is not the configured one.

### 7.3.2 From energy deposition data

An energy deposition file from a particle simulation can be used to generate photons. The file lists deposition nodes, each containing the deposited energy, position, time and the material index. Photons of both the primary and/or secondary scintillation can be generated.

If primary scintillation is enabled, for each deposition node in the file, a number of photons is generated isotropically at the node position. The number of photons is defined by the product of the deposited energy and the photon yield defined for the material at this position, “blurred” by the intrinsic energy resolution of that material (Gaussian distribution). The emission wavelength and the time stamp are defined by the emission spectrum and the rise/decay times of the primary scintillation for that material.

If the secondary scintillation is enabled, the generation process is the following. The number of ionisation electrons is computed based on the deposited energy and the energy per e-/ion pair defined for the material at the position of the node. It is assumed that the electrons drift in the positive direction of the Z global axis, until they reach an object with the special role of *Secondary scintillator*. The drift of each electron is treated individually. The lateral drift is simulated based on the transverse electron diffusion coefficients of the materials along the path. The secondary scintillation photons are generated uniformly along the line defined by the overlap of the global Z axis with the secondary scintillator object, passing the position where the electron enters the secondary scintillator. The number of photons per electron is defined by the corresponding property of the material of the secondary scintillator. The wavelength is sampled from the secondary scintillation spectrum. The time stamp is generated based on the deposition node time shifted by the drift time and assuming the decay time configured for the material of the secondary scintillator.

### 7.3.3 From file

Photons can also be generated based on the individual records read from a file. Each record lists the position, direction, time and waveindex of a photon. The photon records are grouped into events using new event markers. Detailed information on the format is given in the GUI.

## 7.4 Scoring

### 7.4.1 Light sensors

The special role of *light sensor* can be assigned to any physical object defined in the geometry. The sensors simulate the operation of photomultipliers (PMT) or silicon photomultipliers (SiPM) by recording the number of detected photon hits per event. A “sensor model” has to be assigned to each sensor to define the properties related to the photon detection and signal generation, such as the photon detection efficiency, angular and area response, and dark count rate.

A model can be configured to represent a SiPM, and in this case, the number of pixels has to be defined: if multiple photons hit a given pixel during the same event, the pixel can register only one hit. It is possible to convert the number of detected hits into a readout signal (e.g by sampling a Gamma distribution to simulate the single photoelectron spectrum of a PMT) and add electronic noise.

All light sensors (including their copies from copy-generating logical objects in the geometry), receive unique identification indexes. The sensor indexes can be displayed in the Geometry visualisation window, and their coordinates can be requested in script using the *getLightSensorPositions* method of ***geo*** unit. The sensor model properties can be modified in the Sensors window in the GUI or by using the ***sens*** scripting unit.

### 7.4.2 Optical monitors

Optical monitors operate similarly to the particle monitors (see section 6.3.2). The main difference is that they record photon wavelength information instead of the particle energy.

### 7.4.3 Logs and statistics

Acquisition of the following statistics can be activated in a photon simulation: the number of times each of the optical processes has been triggered (e.g., Fresnel reflection or scattering due to an interface rule), the number of specific photon end-of-life processes (such as escaping the World object or absorption) and the light sensor statistics for detected photons. The light sensor data include the distributions of photon wavelength, time, angle of incidence and number of tracking steps before detection.

When it is required to implement a custom detection check for light sensor hits, it is possible to activate sensor hit logging. In that case, when a photon enters a sensor object, a record is saved in the output file which includes the sensor index, local position on the sensor, angle of incidence, time and waveindex.

Ants3 can also generate a simplified or a detailed tracing log. The simplified one contains only the tracing step positions and is mainly intended for the visualization of the photon “tracks”. The detailed log stores all available data for each tracing step, such as the process name, position, time, waveindex and material index. To reduce the size of the output file, logging can be limited by a set of conditions. One can define photon processes which should (or should not) be triggered during tracing of the photon and the objects in the geometry which should (or should not) appear in the photon path.

## 7.5 Running the simulation

After configuring the geometry, materials, photon generators, scoring and, optionally, the interface rules and functional objects, the final step before running a simulation is to set the number of events, the seed of the random generator and the output options. These options include the output directory and the enable/disable flag for each scorer type.

A simulation run can be started using the GUI controls at the *Simulation* tab of the Photon simulation window or by calling the *simulate* method of the ***lsim*** scripting unit. As in particle simulations, Ants3 distributes the workload automatically over the local processes and those on the configured nodes of the farm. The output files from all processes are automatically merged and the resulting files, one per activated scorer type, are saved in the configured output directory. Targeting simulations on a computer cluster, it is also possible to run *lsim* executable directly from the terminal with the name of the configuration json file provided as the argument.

## 7.6 Analysis tools

Ants3 offers a comprehensive system of analysis tools, both in the GUI (the *Results* tab of the Photon simulation window) and in the scripting interface.

### 7.6.1 Sensor signals

The sensor signals can be visualised on an event-by-event basis either as a table or using a dedicated viewer. The viewer shows the sensor positions and shapes (respecting the scale), and the sensor signals are displayed both as the sensor color and a numeric value on top of the corresponding sensor (see Figure 10). It is also possible to limit visualisation to a custom subset of the sensors by indicating their indexes.

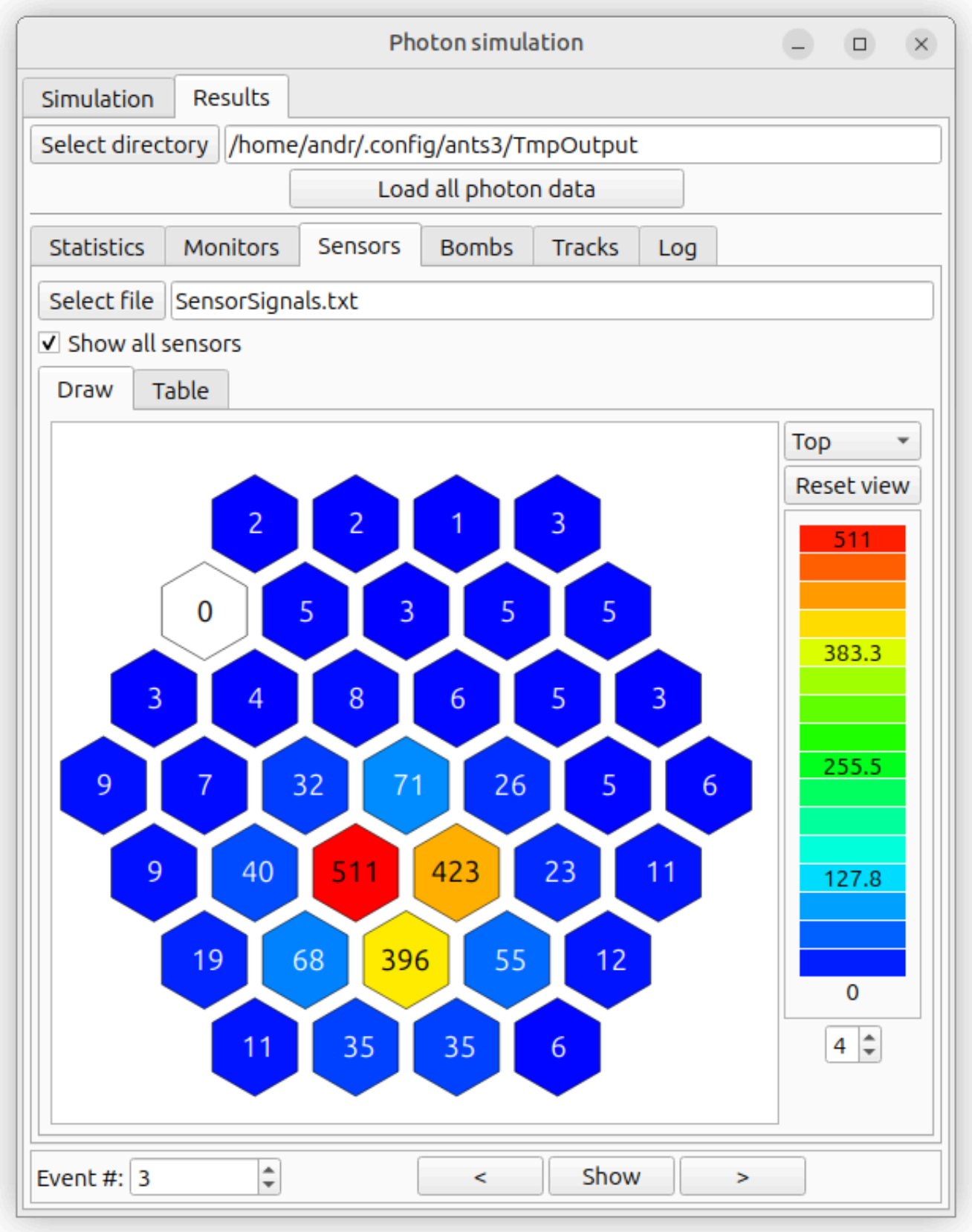

Figure 10. Signal viewer for the light sensors.

### 7.6.2 Monitors

The visualisation and handling of the data recorded by the optical monitors are identical to those of the particle monitors (see section 6.5.2). The only exception is that, instead of the particle energy, the photon wavelength (or, alternatively, the waveindex) is shown. The GUI controls are situated at the *Monitors* tab, and the scripting unit is ***lsim***.

### 7.6.3 Tracks

The tools for the visualisation of photon tracks are given at the *Tracks* tab. The tracks are shown on an event-by-event basis in the Geometry visualisation window. The default colors are red for photons detected by a sensor, and teal / magenta for primary / secondary scintillation, respectively, for the photons that were not detected. It is possible to limit visualization to a selected number of tracks and only to photons detected by a sensor.

### 7.6.4 Photon bombs

The *Bombs* tab gives access to the visualisation of photon bomb positions in the Geometry window, on an event-by-event basis or combined across all events. Note that it is possible to add custom markers to be shown in the same window using ***geowin*** scripting unit, thus providing infrastructure for visual comparison of the true positions of the simulated events with the reconstructed ones, based, for example, on the recorded sensor signals.

### 7.6.5 Statistics

The *Statistics* tab shows the statistical information on the optical processes triggered during tracing and, specifically, gives a detailed break-down on the processes resulting in the photon's end of tracing. The tab also contains controls to plot the distributions of the number of tracing steps, wavelength, time and angle of incidence for the photons detected by the sensors.

### 7.6.6 Photon log

The *Log* tab hosts control for visualisation of the detailed tracing log. The log is shown on a photon-by-photon basis, listing all the tracing steps and including the process name, waveindex, position, geometric object with its copy index and time. The corresponding photon "track" is drawn in the Geometry visualisation window. Filters can be applied to find a photon in the log according to the specific criteria by defining the photon processes which should (or should not) be present and/or the objects which should (or should not) appear in the photon path.

# 8. Visualisation of graphs and histograms

Ants3 gives access to drawing histogram and graph objects of the CERN ROOT package. The plots are shown in the Graph window (see Figure 11), which offers a comprehensive set of tools for the visualisation and analysis of the results.

A plot shown in the window can be stored in the "basket" (see the panel with that name in Figure 11) and it is possible to redraw any plot stored there by clicking the corresponding item. Combined plots can be assembled by drawing one plot from the basket on top of the other. The entire basket can be saved to or loaded from a file.

Controls are offered to change the line and marker properties, zoom, and add a legend or an item to the plot (e.g. a text box or a line). It is possible to generate a "template" for a plot, copying the format of the lines/markers, fonts of the labels/titles, axis ranges, etc, and apply it to any other plot (entirely or only selected features).

Tools are provided for normalization, scaling and shifting plots, fitting the data and making projections for 2D histograms (including projection to an arbitrary line not aligned to the histogram axis). Multiple plots can be shown in a multi-pad view with flexible controls for adjustments of the margins, scaling of the label font and adding identification boxes. The drawn images can be saved to a file using one of the available raster or vector formats or copied to the system clipboard.

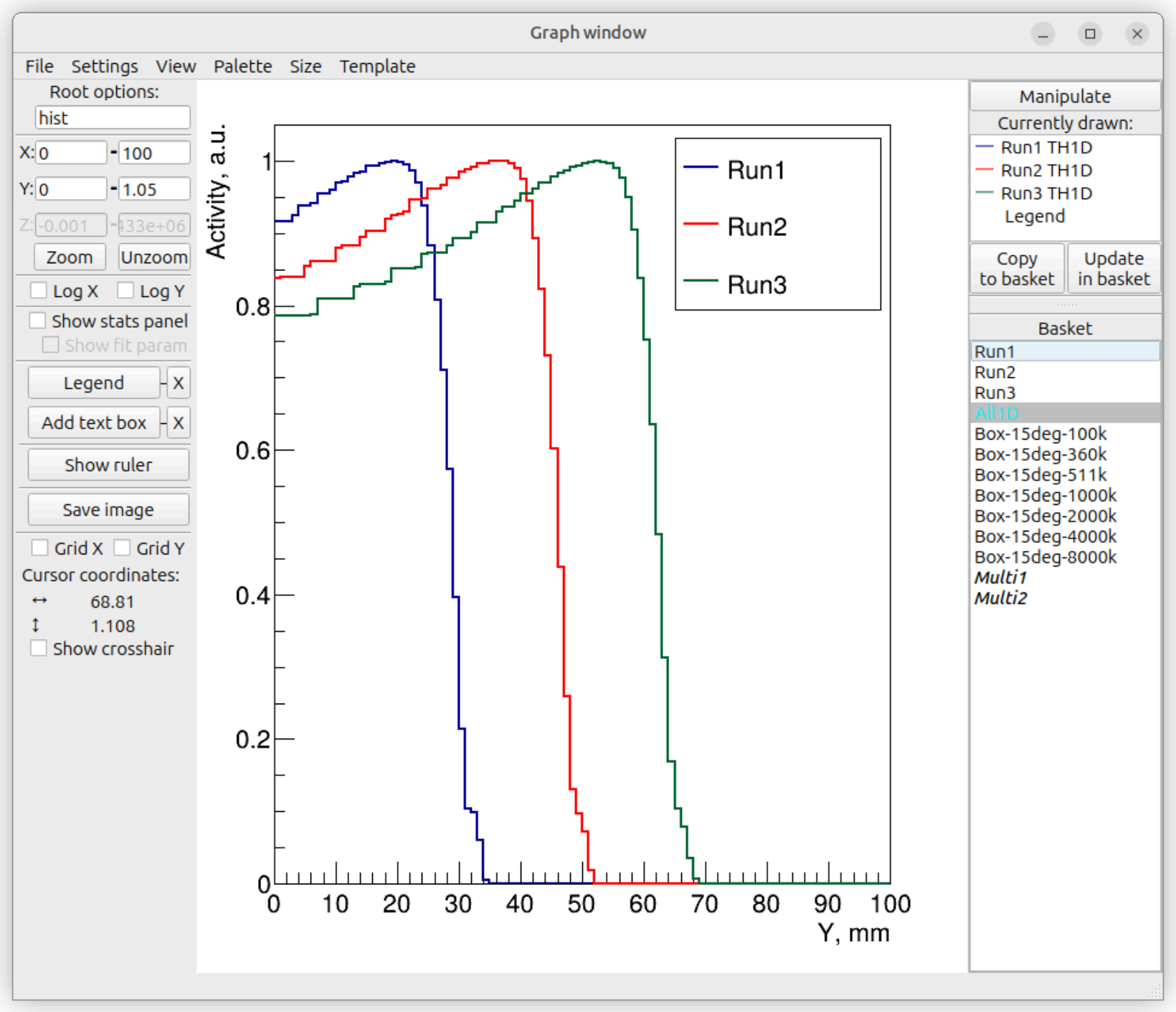

Figure 11. Graph window of Ants3.

# 9. Implementation details

Ants3 is written in c++ (standard c++17), strongly relying on the Qt framework [15] (GUI, scripting system and network) and CERN ROOT library (geometry, data classes and algorithms).

Ants3 was developed for Linux. More specifically, Ubuntu is recommended as the development team is using it exclusively. The code is not compatible with macOS due to ROOT-Qt communications issues in the graphical system, but, based on the experience with the previous installment of the toolkit, it is likely that Ants3 can run on Windows, however, it has not yet been attempted.

The scripting system is implemented based on the same approach for both supported languages, JavaScript and Python, by using the external interpreters and registering the Ants3 scripting units as global objects. Note that adding new c++ methods to an existing scripting unit is straightforward: a “public slot” method added to the unit's class is automatically made accessible in scripting. JavaScript is supported natively in Qt (QJSEngine class), while for Python the system interpreter is used with a custom interface layer developed to access Qt's “meta object” system.

Ants3 is an open source project. The source code and the detailed installation instructions can be found on GitHub [16].

# 10. Conclusions

A new toolkit, offering a comprehensive front-end for particle simulations in Geant4 and a custom simulator for optical photons, is presented. Due to its fully interactive GUI and an extensive scripting system based on the general-purpose scripting languages, Ants3 has a significantly flatter learning curve compared to Geant4 and most of the other existing front-ends.

Ants3 covers the entire simulation cycle, featuring an intuitive approach for configuration of the geometry and simulation conditions, giving the possibility to automatically distribute work over available local and network resources, and offering a suite of versatile tools to conduct analysis of the results. A significant effort was dedicated to integrate a variety of tests directly accessible during the configuration process to reduce the number of iterations required to set-up a simulation. Wherever possible, alternative approaches for common scoring tasks are provided, thus enabling direct cross-checks.

The primary application area of the toolkit is the development of new detectors and readout methods, with a focus on detector optimization. The targeted user group is very broad, from the experts in the field to those who have little experience in simulations and programming.

As the work on Ants3 continues, the development team welcomes feedback and suggestions. The *Issues* tab at the Github page of the project [16] can be used for this purpose. The users are invited to contribute their configuration examples, material files and script code to be integrated in Ants3.

# Acknowledgments

L. Margato acknowledges FCT-Portugal support through contract CEECINST/00106/2018/CP1494/CT0001.